\documentclass[conference]{IEEEtran}

\ifCLASSINFOpdf
\else
\fi
\usepackage{url}
\usepackage{array}
\usepackage{booktabs}
\usepackage{graphicx} 
\usepackage{float} 

\usepackage{tcolorbox}

\usepackage{scalerel}
\usepackage{tikz}
\usetikzlibrary{svg.path}
\usepackage{hyperref} 

\definecolor{orcidlogocol}{HTML}{A6CE39}
\tikzset{
  orcidlogo/.pic={
    \fill[orcidlogocol] svg{M256,128c0,70.7-57.3,128-128,128C57.3,256,0,198.7,0,128C0,57.3,57.3,0,128,0C198.7,0,256,57.3,256,128z};
    \fill[white] svg{M86.3,186.2H70.9V79.1h15.4v48.4V186.2z}
                 svg{M108.9,79.1h41.6c39.6,0,57,28.3,57,53.6c0,27.5-21.5,53.6-56.8,53.6h-41.8V79.1z M124.3,172.4h24.5c34.9,0,42.9-26.5,42.9-39.7c0-21.5-13.7-39.7-43.7-39.7h-23.7V172.4z}
                 svg{M88.7,56.8c0,5.5-4.5,10.1-10.1,10.1c-5.6,0-10.1-4.6-10.1-10.1c0-5.6,4.5-10.1,10.1-10.1C84.2,46.7,88.7,51.3,88.7,56.8z};
  }
}

\newcommand\orcidicon[1]{\href{https://orcid.org/#1}{\mbox{\scalerel*{
\begin{tikzpicture}[yscale=-1,transform shape]
\pic{orcidlogo};
\end{tikzpicture}
}{|}}}}

\usepackage{hyperref}

\hypersetup{
  pdfkeywords={IoT, Internet of Things, security, privacy, Matter, GitHub, developers, interoperability, technical challenges, open-source, connectivity, CSA, Project CHIP, smart home devices, software development, integration, standards, developer experience, empirical analysis, usability issues, technical documentation, community engagement, collaboration, deployment, testing infrastructure, cross-vendor collaboration, software engineering},
  pdfauthor={Muhammad Hassan},
  pdftitle={Insights from GitHub Community on the Matter Standard:
Developer Perspectives and Challenges}
}

\begin{document}
%
\title{Insights from GitHub Community on the Matter Standard:
Developer Perspectives and Challenges}

\author{\IEEEauthorblockN{Muhammad Hassan\IEEEauthorrefmark{1}\orcidicon{0000-0002-5713-9658},
Carl Gunter\IEEEauthorrefmark{1},
Susan Landau\IEEEauthorrefmark{2}, 
Masooda Bashir\IEEEauthorrefmark{1}}
\IEEEauthorblockA{\IEEEauthorrefmark{1}University of Illinois Urbana Champaign\\
Email: \{ \href{mailto:mhassa42@illinois.edu}{mhassa42}, \href{mailto:cgunter@illinois.edu}{cgunter}, \href{mailto:mnb@illinois.edu}{mnb} \}@illinois.edu}
\IEEEauthorblockA{\IEEEauthorrefmark{2}Tufts University  \href{mailto:susan.landau@tufts.edu} {\{susan.landau@tufts.edu\}}} 
}


\IEEEoverridecommandlockouts
\makeatletter\def\@IEEEpubidpullup{6.5\baselineskip}\makeatother
\IEEEpubid{\parbox{\columnwidth}{
		Workshop on Security and Privacy in Standardized IoT (SDIoTSec) 2026 \\
		23 February 2026, San Diego, CA, USA \\
		ISBN 978-1-970672-01-5 \\ 
		https://dx.doi.org/10.14722/sdiotsec.2026.23046 \\   
		www.ndss-symposium.org
}
\hspace{\columnsep}\makebox[\columnwidth]{}}

\maketitle

\begin{abstract}
Matter seeks to resolve long-standing interoperability problems in the Internet of Things (IoT), yet little is known about how developers experience the standard in day-to-day work. This paper examines over 13{,}000 issues from the official Project CHIP GitHub repository to understand the kinds of problems contributors report when implementing and integrating Matter. Using topic modeling and qualitative analysis, we identify four recurring areas of concern—Testing, Interoperability, Development, and Platform \& Network—and describe how they manifest in the evolution of the codebase and tooling. The findings reveal systematic technical and integration challenges and point to concrete opportunities to refine Matter’s test infrastructure, cross-vendor guidance, and documentation as the standard continues to mature.
\end{abstract}

\section{\textbf{Introduction}}
\label{sec:introduction}

The Internet of Things (IoT) has experienced significant growth in recent years, with estimates suggesting that over 75 billion IoT devices will be in use by 2025~\cite{nccoe}. This rapid expansion reflects IoT's integration across diverse domains, such as workplace environments~\cite{mahler2019working}, home management~\cite{li2021motivations}, healthcare delivery~\cite{al2022iot}, sports applications~\cite{navandar2025modernizing}, and educational settings~\cite{madni2022factors}. This proliferation has created a critical challenge that devices from different manufacturers often struggle to communicate seamlessly. The resulting interoperability gaps have generated inefficiencies, increased deployment costs, and widespread user frustration~\cite{noura2019interoperability}. Understanding how practitioners navigate these challenges is important for advancing IoT adoption and reliability.


In response to these interoperability obstacles, the Matter standard (formerly Project CHIP) was announced in 2019, with the first official specification released in 2022~\cite{alexander_2025}. Matter is an open-source, application-layer connectivity standard intended to support communication among smart home devices from different manufacturers, and its development is coordinated by the Connectivity Standards Alliance (CSA)~\cite{belli2024connectivity}. According to CSA reports, the alliance includes over 550 companies and organizations as of 2025, e.g., Apple, Amazon, Google, IKEA, and Huawei~\cite{joseph_2020, connectivity_standards_alliance_2024, hill_2023}. Earlier consortia, such as the Zigbee Alliance and other home-automation efforts, also sought to improve interoperability but were limited by fragmented ecosystems, varying levels of engagement, and constrained device support~\cite{masreliez2021fragmented, aly2018fragmentation, Noura_Interoperability_IOT}. As of recent CSA and industry reports, more than 1{,}200 devices are Matter-certified, and additional products can participate via bridge mechanisms that connect non-Matter devices into Matter-based systems~\cite{matter_1_2_news_release, briodagh_2023, nordic_semiconductor_2024}. This paper examines how developers experience this emerging standard in practice, rather than evaluating its adoption claims.

Despite Matter's growing deployment and substantial industry backing, the practical realities of implementing this emerging standard remain poorly understood. Prior research has demonstrated the value of systematically analyzing developer-reported issues in related technical domains, surfacing recurring pain points that inform the evolution of frameworks and tools~\cite{buhlmann2022developers, Bissyande_got_issues, wu2024comprehensive}. For emerging standards like Matter, such empirical analysis of developers' lived experiences implementing the specification directly is particularly essential. These insights can inform CSA working groups and member companies that already participate in the Project CHIP repository, helping refine specification documentation, prioritize test and tooling improvements, and adjust guidance so that the standard is easier and safer to deploy.

To address this empirical gap, we conducted a systematic analysis of issues reported on the official GitHub repository for Matter (Project CHIP)~\cite{project_chip}. GitHub Issues serve as a valuable data source because the discussions capture organic developer engagement across the full spectrum of use cases. Unlike formal bug reports or vendor surveys, these threads reflect real-world challenges encountered by both professional developers and hobbyists working across diverse application domains~\cite{proma2024visual}, including integration challenges and specification gaps. By analyzing and categorizing these discussions, we identify the most prevalent technical and usability issues, reveal temporal patterns in problem persistence and resolution, and highlight areas where the Matter specification or its sup-porting ecosystem requires refinement.

Our work aims to make the following contributions:
\begin{itemize}
    \item We present, to the best of our knowledge, the first empirical analysis of developer and practitioner-reported issues related to Matter standard implementation, grounded in systematic analysis of GitHub issue discussions.
    
    \item We identify and categorize the most common themes emerging from reported issues, analyze distribution and resolution timelines across issue types, and uncover patterns in community engagement and communication dynamics.
    
    \item We provide actionable recommendations for the CSA, developers, and the broader IoT community based on empirical evidence, informing future improvements to the Matter specification, developer resources, and best practices in IoT interoperability.
\end{itemize}

\section{\textbf{Background}}
\label{sec:background}
\subsection{\textbf{The Matter Standard}}

To address interoperability issues in smart homes and IoT, major industry stakeholders launched Project Connected Home over IP (CHIP) in 2019, later rebranded as Matter and now developed under the Connectivity Standards Alliance (CSA) and maintained in the official Project CHIP \texttt{connectedhomeip} repository~\cite{project_chip}. Matter is an open-source, application-layer protocol intended to unify smart device connectivity across brands and platforms. A notable feature is its support for \emph{bridge devices}, which allow existing non-Matter devices (e.g., Zigbee or Z-Wave) to participate in Matter-based ecosystems~\cite{ice_2025}. This design extends interoperability while preserving legacy deployments, but also introduces additional configuration and reliability challenges for implementers \cite{pham2023enabling, pizzocolo2025standardization}. Matter further incorporates secure commissioning and certificate-based authentication to protect device integrity and user data~\cite{csa_iot_faq}, and has evolved through successive versions (e.g., Matter~1.5) as vendors add new device classes and features~\cite{zegeye2025comparing, silicon_labs_2025}.

\subsection{\textbf{GitHub and Issue Tracking}}

Modern software engineering relies heavily on collaborative platforms that integrate version control, continuous integration and deployment (CI/CD), and project management~\cite{leite2019survey, luz2018building, feng2023understanding}. GitHub has become a central platform of this kind, combining Git-based version control with tooling for code review, automated testing, and workflow automation~\cite{feliciano2015towards, lima2014coding, seker2021open}. With over 100 million developers and 370 million repositories, it is a key focus for open-source infrastructure and standards development~\cite{dohmke_2023, dennis_2023}.

A core GitHub feature is its issue tracking system. GitHub Issues allow contributors to report bugs, request features, document tasks, and discuss design decisions in an organized and searchable form~\cite{fiechter2021visualizing, al2022improving}. \textit{Labels} and issue types help categorize and prioritize reports, while pull requests provide a structured workflow for proposing and reviewing code changes before merging them into the main branch. Together, these mechanisms support asynchronous coordination and integrate closely with CI/CD pipelines, enabling projects to evolve through iterative, peer-reviewed changes.

Prior work shows that GitHub \textit{Issues} discussions often capture both technical and socio-technical concerns, such as documentation gaps, environment error, and development bugs ~\cite{tsay2014let, mumtaz2022preliminary, al2022improving}. For large and evolving projects, the issue tracker becomes a record of recurring pain points and solutions that may not appear in formal documentation or release notes~\cite{bellomo2016got}. In the case of Matter, the official Project CHIP repository on GitHub hosts the reference implementation, documentation, and support channel, and is used to raise questions, report bugs, and review proposed changes with other users including CSA members and participants ~\cite{matter_project}. The presence of issue and pull-request guidelines provides a structured path for feeding developer feedback into the codebase, which can also be discussed further in CSA working groups\cite{project_chip_2025_contributing}. This study uses GitHub Issues from the Project CHIP repository as an empirical window into how developers and practitioners experience Matter in practice, focusing on challenges, such as technical, interoperability, and network-related problems, and how the community works to resolve them over time.

\section{\textbf{Related Work}}
\label{sec:related}

Understanding the challenges that software engineers and technology practitioners encounter is essential for guiding both standard evolution and tooling improvements. While traditional qualitative methods such as interviews and surveys provide valuable depth, they face constraints in emerging and technical domains. Recruitment is often difficult due to the niche expertise required, and non-disclosure agreements frequently limit candid sharing of implementation experiences \cite{wang2024end, rauf2022challenges, madampe2024struggle}. Moreover, interviews and surveys typically capture a temporal snapshot, making it harder to trace how problems emerge, evolve, and resolve over time \cite{mcleod2011qualitative}. As standards like Matter mature within fast-moving ecosystems, methods that can observe longitudinal practitioner discourse become critical for capturing authentic, evolving challenges.

\subsection{\textbf{Mining Developer Discussions on GitHub}}

Open collaboration platforms such as GitHub provide a unique perspective into the collective problem-solving practices of software engineers. Issue tracking preserves both technical details and the social dynamics of development, in particular, documenting the lifecycle of bugs, feature requests, and design debates  \cite{kalliamvakou2016depth, proma2024visual}. This publicly accessible data enables large-scale empirical analysis of real-world development challenges, complementing traditional qualitative methods with longitudinal, naturally occurring records of developers' experience.

Prior research demonstrates how GitHub Issues can reveal patterns in developer behavior and project health. Zhou et al. found that affective communication, such as emoji use, improves participation and accelerates issue resolution, underscoring how community engagement shapes technical outcomes \cite{zhou2024emoji}. Liao et al. used visual analytics to map engagement and responsiveness within open-source projects, showing how community coordination influences project dynamics and maintenance burdens \cite{liao2018exploring}. Bühlmann and Ghafari examined how developers respond to security vulnerabilities, identifying recurring tensions between openness and risk management in collaborative environments \cite{buhlmann2022developers}. Other studies have analyzed sentiment, language complexity, and responsiveness to trace collaboration bottlenecks and communication norms \cite{destefanis2018measuring, kavaler2017perceived}. In adjacent domains, Zhang et al. applied large-scale opinion mining to MLOps discussions, uncovering systematic practitioner pain points and building taxonomies of recurring issues \cite{zhang2025problems}. Collectively, this body of work establishes that mining open developer communities provides a scalable lens on technical, organizational, and usability challenges.

Analyzing developer discussions offers insight into the iterative improvement of standards and their real-world adoption. By examining technical problems and issues, researchers can identify specification ambiguities and limitations, or ecosystem dependencies impact progress. Such insights can feed back into standardization bodies and technical teams, promoting better alignment between specification design and implementation feasibility \cite{sharma2017investigating, mannan2020relationship, sulun2024empirical}. For emerging standards like Matter, where both technical maturity and ecosystem support are still evolving, this feedback mechanism becomes especially important.


\subsection{\textbf{Developer Perspectives on Matter Standard}}

Matter standard for the Internet of Things aims to reduce fragmentation and enable seamless device communication, yet its real-world adoption depends heavily on developer experience. Prior work on Matter standard has largely focused on architectural evaluations, security analyses, or user-centric studies of smart home adoption, leaving the practitioner’s implementation perspective underexplored. Zegeye et al. developed a hardware testbed to evaluate Matter’s technical capabilities, demonstrating that integrating heterogeneous devices remains challenging even under controlled conditions \cite{Zegeye_connected_home}. Mangar et al. designed an experimental Matter deployment to assess user experience, revealing usability frictions that emerge in practice \cite{mangar2024designing}. These studies highlight Matter’s potential but also underscore that specification design and implementation reality can vary.

More recent work has identified privacy and security risks introduced by Matter’s interoperability features, such as vulnerabilities in device pairing and delegation that enable hidden hub eavesdropping attacks \cite{liao2024wip, wang2025hidden}. These findings suggest that the same mechanisms enabling flexible cross-device control can create security pitfalls. However, such analyses typically focus on protocol design rather than on the experiences of developers encountering these issues in production code. Understanding the practitioners and developers' challenges essential for informing both specification refinements and tooling support. By systematically examining how developers themselves report Issues from the Project CHIP repository and resolve such challenges in the Matter codebase, this study 
complements prior research providing empirical insights into the technical, usability, and process challenges that shape Matter’s ongoing development and adoption.


\section{\textbf{Methodology}}
\label{sec:methods}

\begin{table*}[ht] 
    \centering
    \setlength\extrarowheight{3.5pt} 
    \title{Topics \& Keywords}
    \caption{Topic numbers along with labels and categories for generated Matter GitHub Issues.}
    \label{visual:topics_table}
    \resizebox{\textwidth}{!}{ 
    \begin{tabular}{@{\extracolsep{5pt}}l l l l@{}}
        \toprule
        \textbf{Topic\#} & \textbf{Keywords} & \textbf{Label} & \textbf{Category} \\
        \midrule
        0 & chip, info, file, python, lib, line, src, connectedhomeip, test, stdout & Matter Build \& Errors & Development \\
        1 & thread, wifi, ble, matter, network, commissioning, nxp, silabs, platform, support & Network \& Platform Setup & Platform \& Network \\
        2 & build, matter, include, darwin, linux, building, arm, readable, file, export & Build Systems & Development \\
        3 & test, tc, chip, spec, connectedhomeip, cluster, specifications, xml, attribute, python & Specification Testing & Testing \\
        4 & dl, type, dmg, ota, msg, key, em, command, dis, received & Connectivity & Platform \& Network \\
        5 & app, cluster, clusters, chip, server, cpp, attributes, attribute, data, connectedhomeip & Cluster Management & Development \\
        6 & class, testing, support, unit, tests, function, implementation, api, data, functions & Hardware-based Security & Testing \\
        7 & testing, tests, fix, changes, files, test, fixes, zap, summary, unit & Testing Maintenance & Testing \\
        8 & chip, tool, log, dmg, command, dut, tc, error, test, user & Testing Tools & Testing \\
        9 & third, party, pigweed, lwip, source, pw, file, python, src, target & Dependency and Environment Configuration & Interoperability \\
        10 & request, guidelines, pydata, actions, sphinx, theme, formatting, force, readability, apply & Documentation Maintenance & Development \\
        11 & app, esp, android, tv, camera, linux, casting, using, controller, webrtc & Application Interface & Interoperability \\
        12 & platform, response, version, sdk, bug, reproduction, hash, prevalence, tested, type & Cross Platform & Interoperability \\
        13 & openthread, details, summary, close, reopen, merge, br, redirect, upgrade, creating & Third-Party Dependency & Platform \& Network \\
        14 & test, coding, cluster, problem, location, endpoint, spec, conformance, name, rule & Cluster and Testing Conformance & Testing \\
        15 & device, icd, fabric, check, discussion, connectedhomeip, apple, subscription, chip, value & Fabric State Management & Interoperability \\
        16 & connectedhomeip, chip, doc, update, title, testing, size, docker, see, increase & Docker Build Environment & Development \\
        \bottomrule
    \end{tabular}
    }
\end{table*}

\subsection{\textbf{Data Collection and Preparation}}

This study analyzes publicly available issue discussions from the official Matter specification repository, \texttt{project-chip/connectedhomeip}, hosted on GitHub. The dataset was collected using the GitHub REST API and includes all public issues from the repository’s creation in 2020 through August 6, 2025. This time window spans multiple Matter specification releases, including version~1.4, and captures the ecosystem’s evolution over several development cycles. For each issue, we retrieved the title, body, status (open/closed), creation and closure timestamps, labels, and author identifiers.

We then preprocessed the issue text to prepare it for topic modeling and qualitative analysis. Issue titles and bodies were concatenated into a single document per issue and tokenized using NLTK~\cite{nltk_2025}. Tokens were lowercased and cleaned by removing:
\begin{itemize}
    \item Code blocks enclosed in triple backticks and inline code spans (which primarily contained stack traces, build logs, and command-line output),
    \item Numeric sequences longer than three digits,
    \item Non-alphabetic characters and single-character tokens.
\end{itemize}
Whitespace was normalized and stop words were removed using NLTK’s English stop-word list, extended with frequent repository-specific terms (e.g., ``github'', ``script'', ``commit'', ``issue'') to reduce noise and focus on substantive technical content. After preprocessing, the final corpus comprised 13{,}008 issues related to the Matter specification and its reference implementation. We also verified that removing embedded command output and logs did not materially change the topic structure (further details are provided in Appendix~\ref{app:command_output}).


\noindent \textbf{\textit{Ethical \& Limitations.}}
All data was collected from public GitHub using authenticated API calls and respecting rate limits and the analysis is limited to publicly visible data. Although CSA-affiliated contributors participate in the repository and issues follow structured reporting and review workflows, this dataset does not reveal how often, or how systematically, these reports are incorporated into CSA's internal working-group decisions or changes to the Matter specification.

\subsection{\textbf{Topic Modeling}}

To identify latent themes in developer-reported challenges, we applied Latent Dirichlet Allocation (LDA), a widely used unsupervised topic modeling technique for software repository mining~\cite{chauhan2021topic}. LDA was implemented in Gensim and trained on a bag-of-words representation of the preprocessed corpus, with each issue treated as a separate document and represented using a learned vocabulary dictionary.

We tuned the model through a systematic grid search over three hyperparameters: the number of topics $k$, the number of passes $p$, and the number of training iterations $i$. Concretely, we varied:
\begin{itemize}
    \item $k \in \{5, 15, 25, 35, \ldots, 95\}$,
    \item $p \in \{10, 15, 20\}$,
    \item $i \in \{100, 200, \ldots, 1000\}$.
\end{itemize}
For each configuration, we computed the $C_v$ coherence score, which measures semantic relatedness among high-probability terms within topics and has been shown to correlate with human judgments of topic quality. We selected the model with the highest $C_v$ coherence for subsequent qualitative analysis. The optimal configuration used $k = 17$ topics, $p = 20$ passes, and $i = 100$ iterations, yielding a coherence score of 0.59. Table~\ref{visual:topics_table} summarizes the resulting topics, their top keywords, and the higher-level categories used in our analysis.

\subsection{\textbf{Qualitative Thematic Interpretation}}

To move from automatically discovered topics to interpretable developer challenges, we conducted a multi-stage qualitative interpretation process. For each of the 17 topics, we extracted the top 10 terms and sampled two sets of issues: (i) the 15 issues with the highest proportion of that topic, and (ii) a stratified random sample of 15 additional issues associated with the topic. This design balances exposure to prototypical cases with coverage of more diverse instances and served as the primary basis for our qualitative coding and label assignment. In practice, the term lists and 30-issue samples per topic quickly converged on stable, interpretable labels, and additional spot checks beyond the initial samples did not yield new categories or meanings.

Across all topics, this procedure yields 510 closely examined issues on top of the full quantitative analysis of all 13,008 issues, providing both depth and breadth in our interpretation. An initial investigator (Author~1) reviewed the terms and sampled issues for each topic and assigned a descriptive label reflecting the dominant technical, interoperability, or process concern (e.g., build failures, platform setup, cluster conformance). Two independent reviewers (graduate students with prior research experience in cybersecurity and IoT) then examined the assigned labels, topic keywords, and representative issues. They proposed refinements where labels did not adequately capture the observed discussion patterns. Disagreements were resolved through iterative discussion until consensus was reached.

To further validate the thematic structure, three faculty reviewers with backgrounds in IoT systems, security and privacy engineering, and usable security examined the topic labels, supporting examples, and the grouping of topics into broader categories (Testing, Interoperability, Development, and Platform \& Network). They assessed whether the topics were coherent, whether the labels aligned with the underlying issue content, and whether the category structure reflected meaningful distinctions in the Matter ecosystem. This multi-stage process, which combines automated topic modeling with expert review, follows established practice in qualitative software repository mining and is intended to increase the credibility and interpretability of the resulting themes~\cite{haque2020challenges, agrawal2018wrong, yang2016security, bagherzadeh2019going, abdellatif2020challenges}.

Using the validated labels and categories, we then computed aggregate statistics for each broad category, including issue counts, closure rates, unique participants, comment activity, and time-to-close (Table~\ref{tab:issue_activity_summary}). These quantitative summaries provide a complementary view of how different classes of challenges manifest and are addressed over time within the Matter development community.



\section{\textbf{Result}}
\label{sec:result}

\begin{table*}[!ht]
    \centering
    \caption{Summary of Issue Activity Across Broad Categories}
    \small
    \begin{tabular}{@{}lcccccc@{}}
        \toprule
        \textbf{Broad Category} 
        & \textbf{Issue Count} 
        & \textbf{Percentage} 
        & \textbf{Unique Users} 
        & \textbf{Closed Issues (\%)} 
        & \textbf{Avg. Comments} 
        & \textbf{Avg. Time to Close} \\ 
        \midrule
        Testing                  
        & 5,545 
        & 42.63\% 
        & 501 
        & 4,759 (85.83\%) 
        & 3.53 
        & 26 days \\
        
        Interoperability         
        & 3,926 
        & 30.18\% 
        & 612 
        & 3,040 (77.43\%) 
        & 3.14 
        & 46 days \\
        
        Development              
        & 2,087 
        & 16.04\% 
        & 396 
        & 1,781 (85.34\%) 
        & 3.39 
        & 23 days \\
        
        Platform \& Network      
        & 1,450 
        & 11.15\% 
        & 227 
        & 1,341 (92.48\%) 
        & 3.37 
        & 17 days \\
        \bottomrule
    \end{tabular}
    \label{tab:issue_activity_summary}
\end{table*}

\begin{figure}[t]  
    \centering
    \includegraphics[width=0.7\columnwidth]{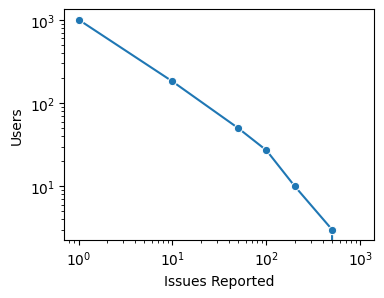}  
    \caption{Number of Users vs Number of Issues Reported}
    \label{fig:num_users_issues}
\end{figure}

\begin{figure*}[t]  
    \centering
    \includegraphics[width=1.9\columnwidth]{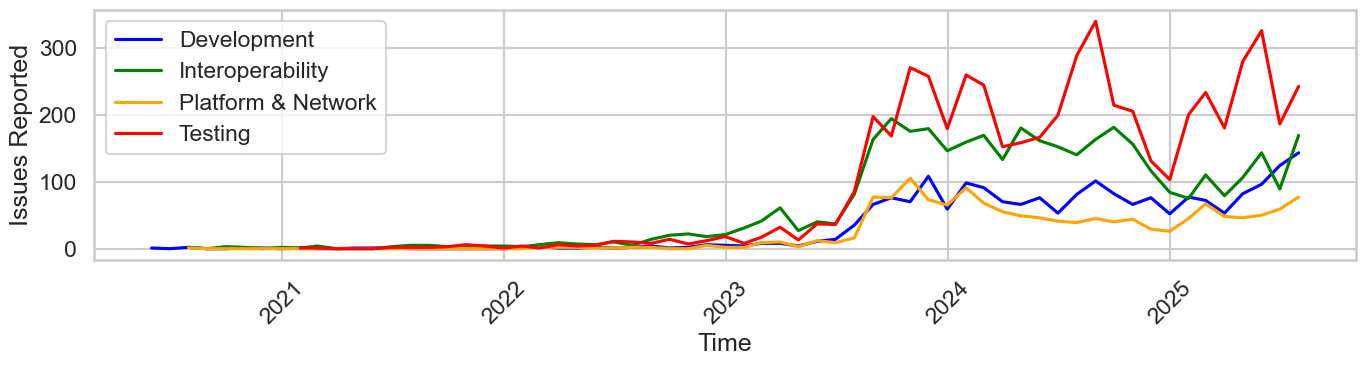}  
    \caption{Number of Issues Reported Over Time by Each Category}
    \label{fig:categories_issues_line}
\end{figure*}

\vspace{-1.2mm}

Our dataset comprises 13{,}008 issues submitted by 1{,}010 unique contributors to the official Matter repository. Figure~\ref{fig:num_users_issues} shows that as the number of issues reported per user increases, the number of users at that reporting level decreases, indicating that a relatively small subset of contributors files a large share of issues\cite{kalliamvakou2016depth, tsay2014let} . On average, a contributor reports approximately 13 issues.

Table~\ref{tab:issue_activity_summary} summarizes activity across the four broad categories derived from topic modeling and qualitative coding. Testing issues constitute the largest group (42.63\%), followed by Interoperability (30.18\%), Development (16.04\%), and Platform \& Network (11.15\%). Each category captures a distinct dimension of practitioner work, ranging from maintaining build and tooling infrastructure to validating cross-platform behavior and network connectivity. Figure~\ref{fig:categories_issues_line} further shows that issue volume grows sharply from late 2022 onward across all categories, corresponding to increased deployment and testing following the initial Matter releases, with Testing and Interoperability dominating the activity over time.

Among all contributors, 288 (about 28\%) self-reported a company affiliation, spanning 166 distinct organizations. These include vendors and platform providers such as Amazon, Apple, Samsung, Nordic Semiconductor, Google, Silicon Labs, and the CSA itself, with Google appearing most frequently (31 users). This mix suggests that both alliance participants and other industry actors use the repository as a shared space for coordinating Matter development.


In the following subsections, we describe each broad category and its topics, and relate them to how developers experience Matter’s implementation, integration, and validation in practice. Table~\ref{tab:term_glossary} in the appendix provides brief explanations of the technical terms used in describing the results.

\subsection{\textbf{Development}}

The Development category captures issues related to building the codebase, maintaining tooling, updating cluster implementations, revising documentation, and managing containerized build environments. This category contains 2{,}087 issues (16.04\%) reported by 396 contributors (39.21\% of all contributors). Two contributors filed more than 100 Development issues. Most issues have been closed (85.34\%), and discussions are active, with an average of 3.39 comments and 3.31 labels per issue. Common labels include \emph{review -- approved}, \emph{review -- pending}, \emph{examples}, and \emph{app}. The average time to close is 23 days, suggesting steady but non-trivial effort to resolve build, tooling, and refactoring tasks. The following describe the five topics covered by \textit{Development} category

\subsubsection{Matter Build \& Errors}


These issues show that build stability is sensitive to rapid dependency changes and cross-language tooling, especially as the project spans multiple operating systems and architectures. Typical reports include build failures, misconfigured scripts, CI pipelines, and compiler errors, often triggered when new features are integrated or shared components are adapted to additional platforms. Examples include failing example applications (e.g., air-purifier conformance failures) and recurring Android and Java build breakages (Issues \#37286, \#34881, \#31509, \#36870).

\subsubsection{Build Systems}


These reports reflect the complexity of sustaining a unified build system for a multi-vendor IoT standard and highlight recurring friction when developers synchronize tooling across environments. Build Systems issues concern the underlying build infrastructure, such as configuration files for build tools (e.g., GN and CMake), branch alignment, and setup scripts for macOS, Linux, Android, and embedded targets. Representative issues involve bootstrap failures, missing Android build artifacts, and version mismatches across release branches (e.g., Issues \#32966, \#31191, \#29935, \#29936).

\subsubsection{Cluster Management}


These reports indicate ongoing evolution of Matter’s data model and the effort required to keep cluster behavior consistent and portable across device types. Cluster Management issues relate to implementing, updating, and refactoring clusters, attributes, and server-side logic. Developers report duplicated callbacks, migrations toward newer abstractions (e.g., \texttt{ServerClusterInterface}), and feature requests for additional cluster types (Issues \#37044, \#38699, \#30234–30235), as well as adjustments to binding behavior and diagnostic reporting.

\subsubsection{Documentation Maintenance}


These reports highlight how documentation directly shapes developer understanding and can either prevent or introduce recurring classes of errors. Documentation Maintenance issues include updates to API descriptions, platform guides, example instructions, and CI/CD documentation. Many issues address formatting problems, version drift, or missing explanations that lead to configuration or build mistakes, such as restyling CI/CD documentation (Issue \#30333) or updating Linux setup instructions (Issue \#32144).

\subsubsection{Docker Build Environment}


These issues highlight the importance of reproducible containerized environments for maintaining a stable build pipeline across diverse vendor toolchains. Docker Build Environment issues focus on container images used in CI and vendor workflows, including updating vendor-specific images, refreshing SDK versions, and addressing resource constraints such as disk usage (e.g., Issues \#29886, \#35569, \#31245, \#33531).


\begin{tcolorbox}
\textbf{\textit{Summary:}} The Development category reveals the ongoing engineering work required to sustain Matter’s shared codebase and tooling. Contributors routinely address build breakages, evolve cluster and server logic, adjust documentation, and maintain Docker-based environments, reflecting the practical cost of supporting a rapidly changing, cross-platform standard.
\end{tcolorbox}

\subsection{\textbf{Interoperability}}

The Interoperability category captures issues that arise when Matter components interact across platforms, applications, third-party dependencies, and device--controller interfaces. It contains 3{,}926 issues (30.18\%) reported by 612 contributors (60.59\% of all contributors). Five contributors filed more than 100 issues in this category, indicating sustained focus on cross-platform and integration-related problems. Most issues have been closed (77.43\%), while 886 issues (22.57\%) remain open. Issue discussions average 3.14 comments and 2.74 labels per issue, with frequent labels including \emph{review -- approved}, \emph{needs triage} (managing and prioritizing issues), \emph{examples}, \emph{darwin}, and \emph{bug}. The average time to close is 46 days, the longest among all categories, reflecting the complexity of diagnosing and resolving integration problems across heterogeneous environments. The Interoperability category contains four topics, each describing a specific set of integration challenges.

\subsubsection{Dependency and Environment Configuration}


These issues underline the difficulty of keeping dependencies aligned across the many platforms that participate in the Matter ecosystem. Dependency and Environment Configuration issues involve setup and maintenance of third-party libraries, build tools, and frameworks. Developers report build failures due to missing or incompatible components, misconfigured Python environments (e.g., SSL errors or missing tools), and version skew in large dependency trees (e.g., Pigweed and Abseil) across Linux, Android, ESP32, NXP, Raspberry~Pi, and other platforms (Issues \#38984, \#29842, \#30634, \#39229).

\subsubsection{Application Interface}


Collectively, these issues show that application-level interoperability depends not only on the specification but also on convergence across platform-specific implementations. Application Interface issues involve how Matter applications behave on different platforms, such as Android, Linux, ESP32, TV casting apps, and camera controllers. Common problems include commissioning-flow failures, missing or inconsistent media features, and mismatched data formats, such as passcode flow differences in TV applications or missing fields in Android JSON encodings (e.g., Issues \#32958, \#38406, \#39152). Reports also highlight gaps in WebRTC-based communication.

\subsubsection{Cross Platform}


These reports emphasize the challenge of maintaining consistent Matter behavior as vendors integrate the standard into their own SDKs and hardware. Cross Platform issues focus on platform-dependent defects, API inconsistencies, and missing feature parity across operating systems. Examples include unsupported commands, missing attributes, performance anomalies on specific hardware, questions about 32-bit Linux support, and attribute type mismatches between Android and iOS (e.g., Issues \#23206, \#22753, \#33400).

\subsubsection{Fabric State Management}


This topic reflects the complexity of ensuring coherent fabric behavior across controllers and devices, particularly when handling long-lived subscriptions and recovery after network disruptions. Fabric State Management issues address fabric data handling, subscription state, idle or check-in behavior, session resumption, and controller--device synchronization. Many reports arise from interactions between the interaction model, subscription lifecycle, and persistent storage, such as implementing and testing check-in counters, resolving subscription resumption failures, and clarifying error handling and metadata representation (e.g., Issues \#30705, \#28903, \#30965, \#17227).

\begin{tcolorbox}
\textbf{\textit{Summary:}} The Interoperability category highlights the effort required to align Matter’s specification with real deployments across platforms, dependencies, applications, and fabrics. The long resolution times and large contributor base indicate that cross-platform behavior and state management are persistent sources of complexity for practitioners.
\end{tcolorbox}

\subsection{\textbf{Platform \& Network Category}}

The Platform \& Network category covers issues related to network configuration, commissioning, and integration with platform-specific networking stacks. It includes 1{,}450 issues (11.15\%) reported by 227 contributors (22.48\%). One contributor filed more than 100 issues. This category has the highest closure rate (92.48\%), with 109 issues (7.52\%) still open. Issues average 3.37 comments and 3.73 labels, with common labels such as \emph{platform}, \emph{review -- approved}, \emph{review -- pending}, \emph{examples}, and \emph{submodules}. The oldest open issue in this category dates to early 2022 and concerns commissioning Ethernet devices over BLE (Bluetooth Low Energy). Overall, these characteristics suggest continuous platform enablement and iterative refinement of network-related components.

\subsubsection{Network \& Platform Setup}


These issues illustrate the ongoing work required to maintain network support across diverse hardware vendors. Network \& Platform Setup issues involve Wi-Fi, Thread, BLE, and commissioning workflows on specific boards and SDKs. Many reports describe platform bring-up and configuration tasks, such as addressing low-power or sleep behavior, enabling support for particular development boards, updating vendor SDKs, or fixing Thread initialization on embedded devices (e.g., Issues \#38078, \#29070, \#37810).

\subsubsection{Connectivity}


Connectivity issues show how bugs in communication or state management can disrupt device behavior across heterogeneous runtime environments. These issues focus on how devices exchange messages and maintain state, including OTA updates, command handling, and subscription behavior. Developers report unintended commands emitted by bridge applications, timeouts during remote updates, firmware update failures on specific boards, problems with subscription resumption, and initialization bugs, as well as runtime issues such as JNI reference leaks (e.g., Issues \#38086, \#28517, \#27460).

\subsubsection {Third-Party Dependency}


Third-Party Dependency issues reflect continuous effort to keep Matter aligned with upstream network stacks and utility libraries while avoiding regressions. These issues capture updates to networking and cryptographic components such as OpenThread, OT-BR-Posix, Pigweed, mbedTLS, nanopb, and tracing libraries. Many issues are triggered by upstream version changes but are critical for platform stability because Matter implementations rely heavily on these dependencies (e.g., Issues \#28774, \#40221, \#29523).

\begin{tcolorbox}
\textbf{\textit{Summary:}} The Platform \& Network category features the essential work of supporting multiple hardware platforms and networking technologies. High closure rates and frequent updates indicate that developers actively maintain low-level networking components and vendor integrations that underpin higher-level Matter functionality.
\end{tcolorbox}

\subsection{\textbf{Testing Category}}

The Testing category constitutes the largest portion of the dataset, with 5{,}545 issues (42.63\%) reported by 501 contributors (49.60\%). Eleven contributors filed more than 100 issues. Most Testing issues are closed (85.83\%), while 786 (14.17\%) remain open. Discussions average 3.53 comments and 3.50 labels per issue, with common labels including \emph{review -- approved}, \emph{tests}, \emph{app}, and \emph{review -- pending}. The average time to close is 26 days. The oldest open issue (~\#4397) dates to early 2021 and concerns separating build components to support independent unit testing. These statistics highlight the centrality of testing in Matter’s development workflow.

\subsubsection{Specification Testing}


Specification Testing issues illustrate continuous alignment work between the evolving Matter specification and the test suite that validates conformance. These issues relate to test-case definitions, XML specifications, cluster attributes, and Python-based test scripts. Many reports document updates required after specification or test-plan revisions, such as adjusting attribute checks, updating XML definitions, or adding new tariff or cluster tests (e.g., Issues \#37988, \#38404, \#34517, \#32937), as well as fixing gaps between YAML/XML specifications and implementations.

\subsubsection{Hardware-based Security}

Hardware-based Security issues show how security-related behavior is incrementally tested and stabilized across platforms. These reports concern unit tests, API refactoring, and behavior related to security and hardware-backed cryptography. Examples include adding or refining security support on specific platforms, restructuring interfaces to improve testability, strengthening integration tests for critical subsystems, and fixing low-level correctness problems such as iterator invalidation or brittle test fixtures (e.g., Issues \#37977, \#36809, \#38008).

\subsubsection{Testing Maintenance}


Testing Maintenance tasks collectively ensure that the testing environment remains consistent with current tooling and code structure. These issues cover upkeep of the test infrastructure, such as fixing compiler warnings, updating ZAP-generated files, regenerating cluster definitions, adjusting CI behavior, and correcting file permissions (e.g., Issues \#39900, \#38820, \#37360).

\subsubsection{Testing Tools}


Testing Tools issues highlight how limitations and bugs in testing tools interact with device behavior to surface failures that might not appear in simpler scenarios. These focus on test execution tools and harnesses, failure traces, and CHIPTool behavior. Many reports describe certification test failures, timeouts, and commissioning reliability problems, particularly under stress tests, as well as failures in semi-automated workflows, trusted-root mismatches, and cluster-specific conformance breakdowns (e.g., Issues \#31467, \#37655, \#33994).

\subsubsection{Cluster and Testing Conformance.}

This topic indicates that application examples often require refinement to match current cluster specifications and conformance rules. These issues arise when example applications fail conformance tests or do not satisfy cluster requirements. Reports describe systematic failures across example apps, such as all-clusters, air-quality-sensor, lock, lighting, and thermostat applications (e.g., Issues \#37269, \#35680, \#37295, \#36824, \#30321, \#30323).

\begin{tcolorbox}
\textbf{\textit{Summary:}} Overall, the Testing category reflects the extensive effort required to maintain compliance with evolving specifications, ensure coverage across hardware and platform environments, and support reliable use of test tools. The size of this category, as shown in Table~\ref{tab:issue_activity_summary}, highlights its central role in Matter’s development and validation workflow.
\end{tcolorbox}

\section{\textbf{Discussion}}
\label{sec:discuss}
\subsection{\textbf{Security- and Privacy-Relevant Issues}}

Security and privacy concerns appear throughout the repository. We identified 725 issues (5.57\% of all issues) containing security or privacy-related keywords, of which 166 remain open. The earliest unresolved issue dates to February 2021 and discusses unsafe API usage. Most of these issues fall within the \textit{Testing} category, reflecting the central role of conformance suites and test harnesses in surfacing security-relevant behavior. The issues cover incorrect privilege checks, insufficient validation of message fields, inconsistent error handling, and missing boundary conditions.


The persistence of unresolved security-related issues over multiple years suggests that it is an ongoing and non-trivial task in a Matter standard as it spans multiple dependencies, programming stacks, and device classes. These results motivate continued effort towards automated fuzzing, continuous conformance testing, and, where feasible, more formal validation of critical paths, particularly around commissioning, access control, and long-lived device subscriptions~\cite{buhlmann2022developers, yang2016security}

\subsection{\textbf{Limits of Developer-Centric Data}}

The Matter repository functions primarily as a coordination space for developers and engineers. Issues are used to track design and development, platform integration, network configuration, test infrastructure, interoperability failures, and dependency maintenance, and to negotiate clarifications when specification text and implementation diverge. This provides a view of implementation-level challenges and the engineering effort required to sustain a large, cross-vendor codebase, however, the dataset does not show how often, or how systematically, these reports feed into CSA working‑group meeting or changes to the Matter specification updates.

Similarly, the repository only partially reflects end-user experience. Pain points such as home deployment difficulties, device onboarding failures, or multi-vendor configuration problems often surface instead in public discussion forums and consumer support communities. As an extension of this work, we plan to combine our corpus with data from user-focused venues, such as Matter Protocol and IoT discussion forums (e.g., Reddit ). This would connect developer-facing issues with the end-user challenges and can help identify where changes to tooling, documentation, or platform defaults would most directly benefit both practitioners and end users.

\subsection{\textbf{Implications for Matter and IoT Standards}}
This study provides an empirical view of how developers experience the Matter standard in practice. To the best of our knowledge, it is the first systematic analysis of Matter-related issues based on a comprehensive examination of the official repository. The results show how contributors use issues to report implementation problems, track issues, and coordinate fixes for the standard.

Our four broad categories, Testing, Interoperability, Development, and Platform \& Network, reveal distinct patterns in both their numbers and resolution behavior. Testing and Interoperability dominate issue counts, and Interoperability issues remain open for longer on average, which suggests that cross-vendor and cross-platform behavior is harder to reproduce, attribute, and resolve than localized build or tooling defects. This pattern is consistent with Matter’s goal of unifying heterogeneous ecosystems and illustrates some of the practical effort required to achieve that goal.

These challenges also point to concrete directions for improvement. The large number of test-related issues indicates that strengthening test frameworks, tooling, and documentation could reduce recurring failures and improve the efficiency of day-to-day development. The amount and duration of interoperability issues suggest a need for clearer cross-vendor guidelines, richer diagnostics, and more comprehensive reference examples for common deployment scenarios. The consistent security issues in testing and commissioning processes highlight the need for automated and formal validation methods for critical protocol behaviors. Together, these findings show that using developer reported GitHub issues as a valuable feedback source for the Connectivity Standards Alliance, implementers and IoT vendors seeking to refine Matter and related interoperability standards.





\vspace{-2mm}

\section{\textbf{Conclusion}}
This study offers an empirical view of how developers engage with the Matter standard through the official Project CHIP repository. Most reported issues concern testing and interoperability, with integration and network-related bugs often proving difficult to resolve. Development and platform support also demand ongoing maintenance, and security-relevant problems continue to appear across versions. Taken together, these findings indicate that improving Matter standard requires continued effort in testing infrastructure, clearer guidance for cross-vendor deployments, and more automated validation of critical protocol workflows as the standard continues to evolve.


\section*{Acknowledgment}

The authors would like to thank Ramazan Yener and Nicholas Yeung for their help and feedback in analysis. This work was supported in part by the National Science Foundation (NSF) under grants 1955228 and 1955805.

\bibliographystyle{plain}
\bibliography{biblio}

\section{\textbf{Appendix}}
\label{sec:appendix}

\subsection{\textbf{Impact of Command Output Removal}}
\label{app:command_output}

While logs and stack traces often contain concrete failure details, we found that including them primarily inflated vocabulary size and emphasized low-level error messages rather than higher-level themes. To assess whether removing command output and logs affected topic structure, we trained LDA models on corpora with and without this content. The best models differed only slightly in topic count (18 vs.\ 17), and their topic–term distributions mapped to the same higher-level categories in qualitative analysis. This suggests that excluding verbose diagnostic log improves semantic coherence without losing important themes. We therefore use the filtered corpus in the main analysis to improve semantic coherence without losing important themes.

\begin{table}[h]
    \centering
    \small
    \caption{Popular User Labels Issue and Their Counts}
    \begin{tabular}{@{}ll@{}}
        \toprule
        \textbf{Category} & \textbf{Count} \\ \midrule
        review - approved & 4708 \\
        review - pending  & 3126 \\
        examples          & 2930 \\
        app               & 2427 \\
        needs triage      & 2168 \\
        fast track        & 2104 \\
        tests             & 1965 \\
        platform          & 1859 \\
        darwin            & 1771 \\
        bug               & 1500 \\
        scripts           & 1363 \\
        controller        & 1271 \\
        github            & 1033 \\ 
        \bottomrule
    \end{tabular}
    \label{tab:user_labels}
\end{table}

\begin{table*}[!t]
\centering
\small
\caption{Selected Technical Terms and Brief Descriptions}
\label{tab:term_glossary}
\begin{tabular}{p{3cm}p{10.2cm}}
\toprule
\textbf{Term} & \textbf{Brief Description} \\
\midrule
Cluster & A functional module in Matter that organizes related commands and attributes (e.g., On/Off, Thermostat). \\
Attribute & A field within a cluster representing the device state or configuration (e.g., current temperature). \\
ServerClusterInterface & An abstraction used in Matter to define server-side behavior for a cluster. \\
CI/CD & Continuous Integration and Deployment; a practice of automating code changes to ensure quality through building and testing. \\
GN & A meta-build tool used to generate project files for native build systems. \\
CMake & A cross-platform tool for configuring and generating build files for C/C++ projects. \\
Docker & A platform that allows the creation of consistent and replicable environments for building and testing applications. \\
Pigweed & A collection of software libraries and tools used in Matter and related projects for embedded systems. \\
Abseil & A set of C++ libraries (e.g., utilities, containers) that serve as dependencies in projects. \\
OpenThread & An open-source implementation of the Thread networking protocol for IoT devices. \\
OT-BR-Posix & An implementation of the OpenThread Border Router designed for POSIX-like systems. \\
mbedTLS & A lightweight cryptographic library used for secure communication protocols like TLS. \\
nanopb & A small-footprint implementation of Protocol Buffers for use in embedded systems. \\
WebRTC & A technology enabling real-time communication for audio, video, and data over the internet. \\
CHIPTool & A reference application for controlling Matter devices, used for testing and commissioning. \\
ICD & Idle or sleep-capable device concept in Matter, with specific behavior for communication and data subscription. \\
JNI & Java Native Interface; a framework that allows Java code to call native applications and libraries written in other languages. \\
SDK & Software Development Kit; a collection of tools, libraries, and documentation for building applications for a specific platform. \\
BLE & Bluetooth Low Energy; a wireless technology designed for short-range communication with minimal power consumption. \\
Fabric in Matter & A concept representing a group of devices or users that share access to a common set of resources in Matter. \\
XML & Extensible Markup Language; a flexible format for structuring data, commonly used in configurations and data exchange. \\
YAML & Yet Another Markup Language; a human-readable data serialization format often used for configuration files. \\
\bottomrule
\end{tabular}
\end{table*}


%



\end{document}